\documentclass[aps,prl,twocolumn, nofootinbib, preprintnumbers, english]{revtex4-1}
\usepackage{amsthm}
\usepackage{amsmath}
\usepackage{amsfonts}
\usepackage{amssymb}
\usepackage{graphicx}
\usepackage{babel}
\usepackage{verbatim}

\newcommand{\ddbar}{\mathrm{d}\hspace{-9pt} -}
\newcommand{\bra}[1]{\left\langle #1\right|}
\newcommand{\ket}[1]{\left| #1\right\rangle}

\begin{document}
\title{Loop current order and $d$-wave superconductivity: some observable consequences}
\author{Andrea Allais, and T. Senthil}
\affiliation{Department of Physics, Massachusetts Institute of Technology,
Cambridge, MA 02139, USA }
 
\date{\today}
\begin{abstract}
Loop current order has been reported in the pseudogap regime of a few cuprate systems in polarized neutron scattering experiments. Here we study several observable consequences of such order in the $d$-wave superconducting state at low $T$. The symmetries of the loop order removes degeneracy between momenta $\vec k$ and $-\vec k$. Consequently there is a remnant Bogoliubov Fermi surface in the superconducting state. Bounds on the possible existence of such a Fermi surface may be placed from existing data. Detecting  such a Fermi surface will be a very useful confirmation of the existence of loop order in various cuprates. We show through explicit calculations that the Bogoliubov Fermi surface does not display quantum oscillations in a magnetic field consistent with natural expectations. Inclusion of a field induced spin stripe order reconstructs the Bogoliubov Fermi surface to develop pockets which then show quantum oscillations in the superconducting state. Difficulties with interpreting quantum oscillation data in the cuprates along these lines are pointed out.    

 \end{abstract}

\maketitle 

\section{Introduction}

In the last few years there have been several reports in polarized neutron scattering experiments of a time reversal breaking magnetic order in the pseudogap regime  of the underdoped cuprates \cite{PhysRevLett.96.197001, PhysRevB.78.020506, li_unusual_2008}. The order does not break the lattice translational symmetry, and the unit cell does not have a net magnetic dipole moment, so it is due to higher moments. These observations are consistent with a proposal by Varma \cite{PhysRevB.73.155113, PhysRevLett.89.247003, PhysRevLett.83.3538}, that, in the pseudogap phase, the unit cell carries permanent circulating current loops. Much remains unclear theoretically on the origin of this loop current order as well as its implications for understanding other aspects of cuprate phenomenology. In this paper we describe several observable properties of the presence of loop current order in the {\em superconducting} state at low temperature. 

Most of the experimental evidence for the presence of loop current order is currently restricted to temperatures above the SC transition. Nevertheless we shall assume that the order coexists with SC at low-$T$ and explore the properties of the resultant state. As discussed earlier by Berg et al\cite{berg_stability_2007}, the loop current order profoundly modifies the gapless fermionic excitations of the $d$-wave superconductor with immediate implication for the low-$T$ thermodynamics and for photoemission spectra. Specifically the gapless nodal points of the ordinary $d$-wave state are converted to gapless Fermi surfaces of Bogoliubov quasiparticles. Existing data in the superconducting state may then be used to put bounds on the possible existence of such a Bogoliubov Fermi surface. Observation of the corresponding modified low energy properties in careful future experiments will thus be significant evidence for the occurrence of loop current order in the low temperature SC state. 

The presence of a Fermi surface of Bogoliubov quasiparticles immediately leads to the question of whether this SC state may support quantum oscillations. We demonstrate through explicit calculations that it does not. The underlying reason is that the superposition of electron and hole excitations which makes up the Bogoliubov quasiparticle changes character as we go around the Fermi surface. On one portion it mainly has electron character which changes to hole character on a different portion. Consequently it is natural that there is no Landau level formation of the low energy states. In the cuprates, quantum oscillations are observed to set in at high field in the mixed state \cite{doiron-leyraud_quantum_2007, PhysRevLett.100.047004, PhysRevLett.100.047003, PhysRevLett.100.187005, PhysRevB.81.140505, sebastian_towards_2011, riggs_heat_2011}. Further there is strong evidence for the onset of translation symmetry breaking density wave order in a field \cite{lake_spins_2001, PhysRevB.71.220508, PhysRevLett.102.177006, PhysRevLett.103.017001, wu_magnetic-field-induced_2011, laliberte_fermi-surface_2011}, and it is clearly necessary to incorporate this order to discuss the quantum oscillations.  For wavevectors appropriate to the cuprates we show that spin density wave order reconstructs the Bogoliubov Fermi surface to produce a pocket where the quasiparticle has mostly electron (or hole) character through out the Fermi surface. Consequently this pocket may be expected to show quantum oscillations. Our explicit calculations confirm this expectation.

We critically evaluate the question of whether the high field state studied in the quantum oscillation experiments could be understood as a $d$-wave superconductor coexisting with both loop current and stripe order, and point out several difficulties with this idea. In particular the strength of loop order required to reproduce the frequency of the oscillation seems to be much higher than is allowed by constraints coming from zero field experiments.

Much of the theoretical discussion of loop currents in the cuprates has emphasized the apparent importance of multiple bands. However as pointed out in Ref. \onlinecite{PhysRevB.69.245104}, a single band model already can support a loop current pattern that is identical in symmetry to the popular current pattern of Varma. We point out here that in the standard single band model of hole doped cuprates with a first neighbor hopping $t$ and second neighbor hopping $t'$, the ratio $t'/t$ is negative. This means that an elementary triangular loop inside the unit cell is frustrated and this may provide some energetic gain in stabilizing the loop pattern. For the electron-doped cuprates $t'/t > 0$ and the absence of frustration suggests that the loop order may also be absent. It will thus be interesting to search for loop order in the electron doped cuprates. 

\section{Bogoliubov Fermi surface in the superconductor}

We will phrase our discussion in terms of the usual one band model though our results depend only on the symmetry of the state and should thus hold more generally.  Deep in the superconducting state the quasiparticle dispersion is expected to be modeled well by a quadratic Bogoliubov Hamiltonian: 
\begin{equation}
H = \sum_{k}\left[ \epsilon_k \sum_\sigma c^\dagger_{k \sigma} c_{k \sigma} + \Delta_k c^\dagger_{k\uparrow} c^\dagger_{-k \downarrow} + \text{h.c.}\right]\,.
\end{equation}
We take $\Delta_k \sim 2\Delta_0\left(\cos k_x - \cos k_y\right)$. The `normal' state dispersion is taken to be that of a tight binding model with nearest and second neighbor hopping which incorporates the loop current pattern. 
\begin{equation}
H_0 =-\sum_{x,\sigma,\nu}t_{\nu}c_{x+\nu,\sigma}^{\dagger}c_{x,\sigma}\,.
\end{equation}
Here $\nu$ labels the neighbors, with $t_{-\nu}=t_{\nu}^{\star}$, and, as a special case, $t_{0}=\mu$. The current operator in this hopping Hamiltonian is given by the usual expression
\begin{equation}
j_{\nu}(x)=i\sum_{\sigma}(t_{\nu}c_{x+\nu,\sigma}^{\dagger}c_{x,\sigma}-\text{h.c.})\,.
\end{equation}

\begin{center}
  \includegraphics{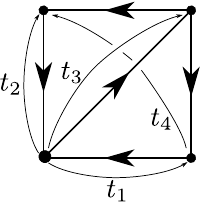}
\end{center}
In our case, we choose to have, with reference to the diagram above, $t_{1}$, $t_{2}$ and $t_{4}$ real, and to give an imaginary part to $t_{3}$. This choice yields the current pattern shown. In addition to time reversal, this pattern breaks reflection about $\hat{x}$,$\hat{y}$ and $\hat{x}+\hat{y}$, rotations of $\pi/2$ about a site, but preserves reflection about $\hat{x}-\hat{y}$. The expectation value of the loop current (in the ``normal" state) is given by
\begin{equation}\label{current}
 \left\langle j\right\rangle = 4 t_1 \int \ddbar^2\! k\, \theta\left(-\epsilon_k\right) \sin k_x\,.
\end{equation} 

The simultaneous breaking of both time reversal and inversion implies that  $\epsilon_{k} \neq \epsilon_{-k}$. Then the superconducting pairing does not fully gap out the Fermi surface: close to the nodes, where the superconducting order parameter is small, there remain pockets of gapless excitations\cite{berg_stability_2007}. To be more definite, the spectrum of the Bogoliubov quasiparticles is
\begin{equation}
\begin{aligned}
&E^\pm_k = \epsilon_{k}^{a}\pm E_k^0\,, && E_k^0 = \sqrt{\left(\epsilon_{k}^{s}\right)^{2}+|\Delta_{k}|^{2}}\,,
\end{aligned}
\end{equation}
where
\begin{equation}
\epsilon_{k}^{s}=\frac{\epsilon_{k}+\epsilon_{-k}}{2}\,,\quad\epsilon_{k}^{a}=\frac{\epsilon_{k}-\epsilon_{-k}}{2}\,.
\end{equation}

\begin{figure}
\begin{center}
 \includegraphics[width=0.5\columnwidth]{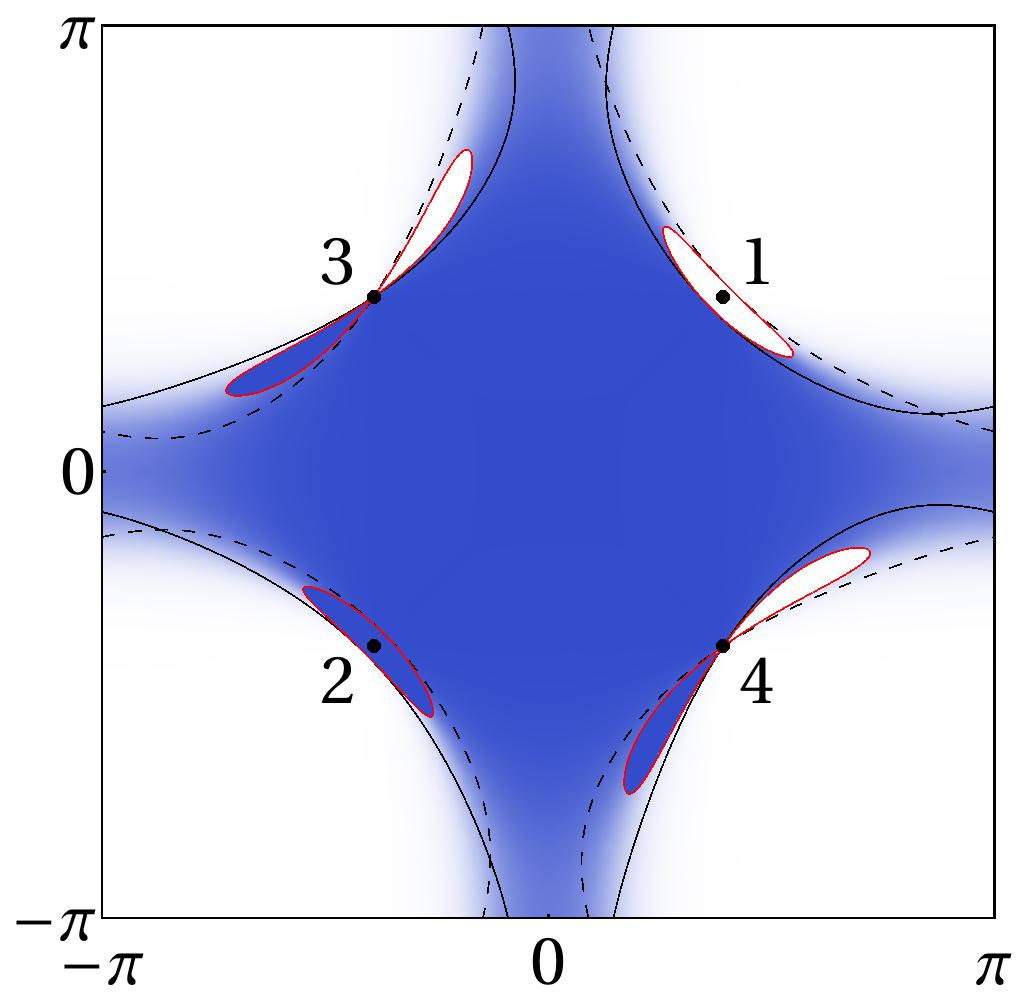}%
 \includegraphics[width=0.5\columnwidth]{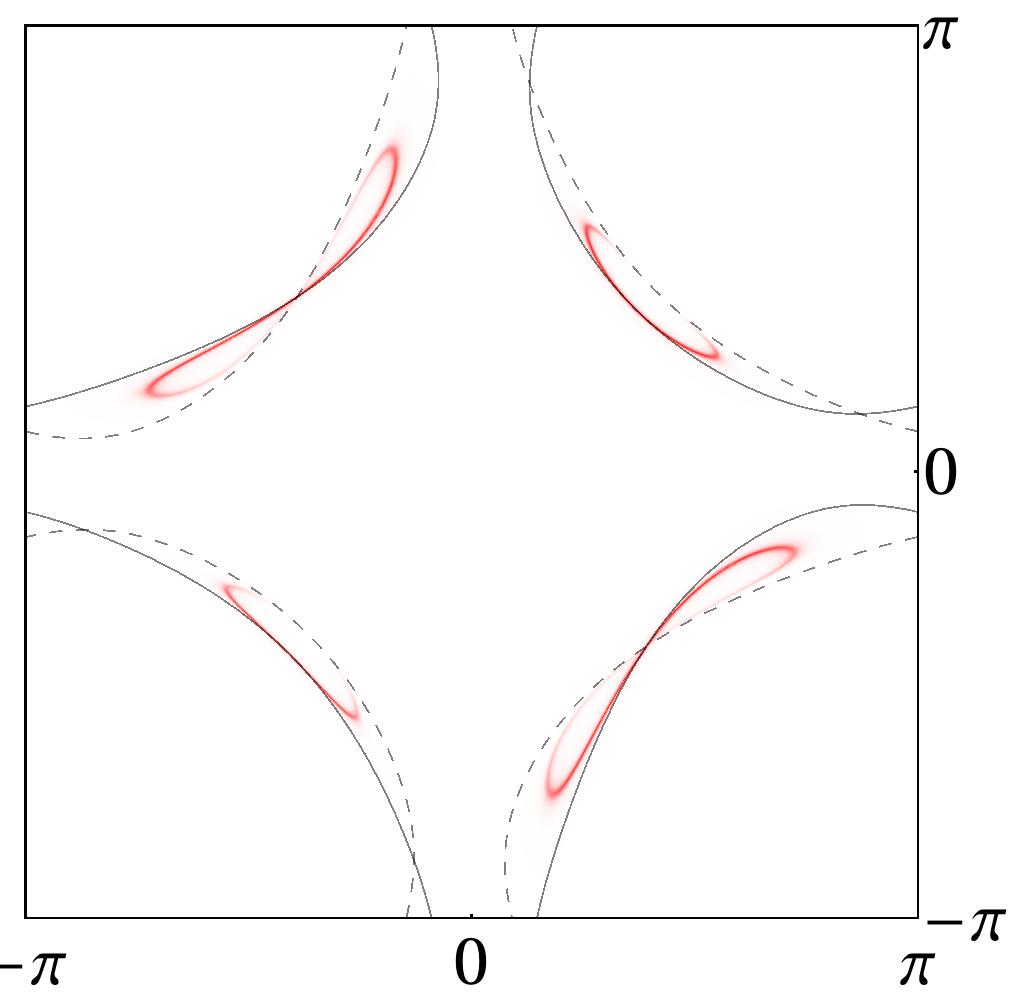}

 \includegraphics[width=0.5\columnwidth]{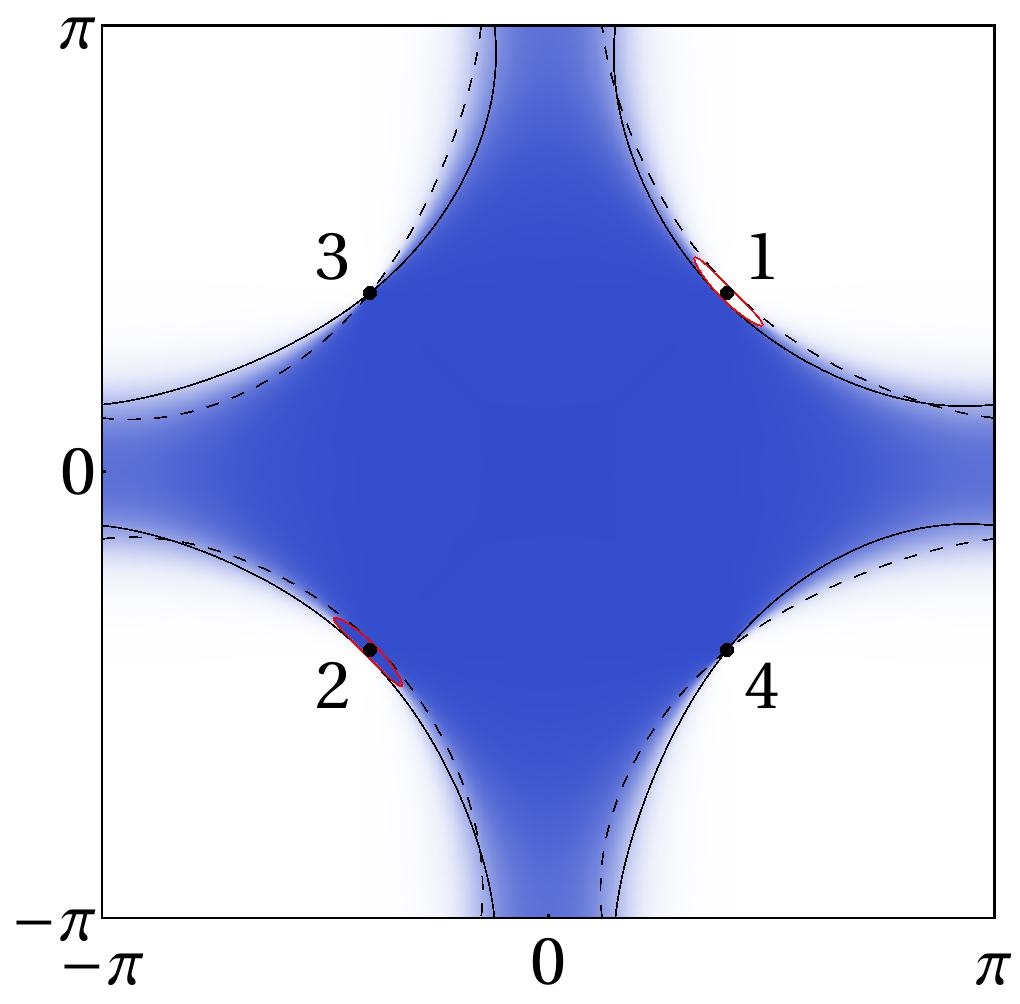}%
 \includegraphics[width=0.5\columnwidth]{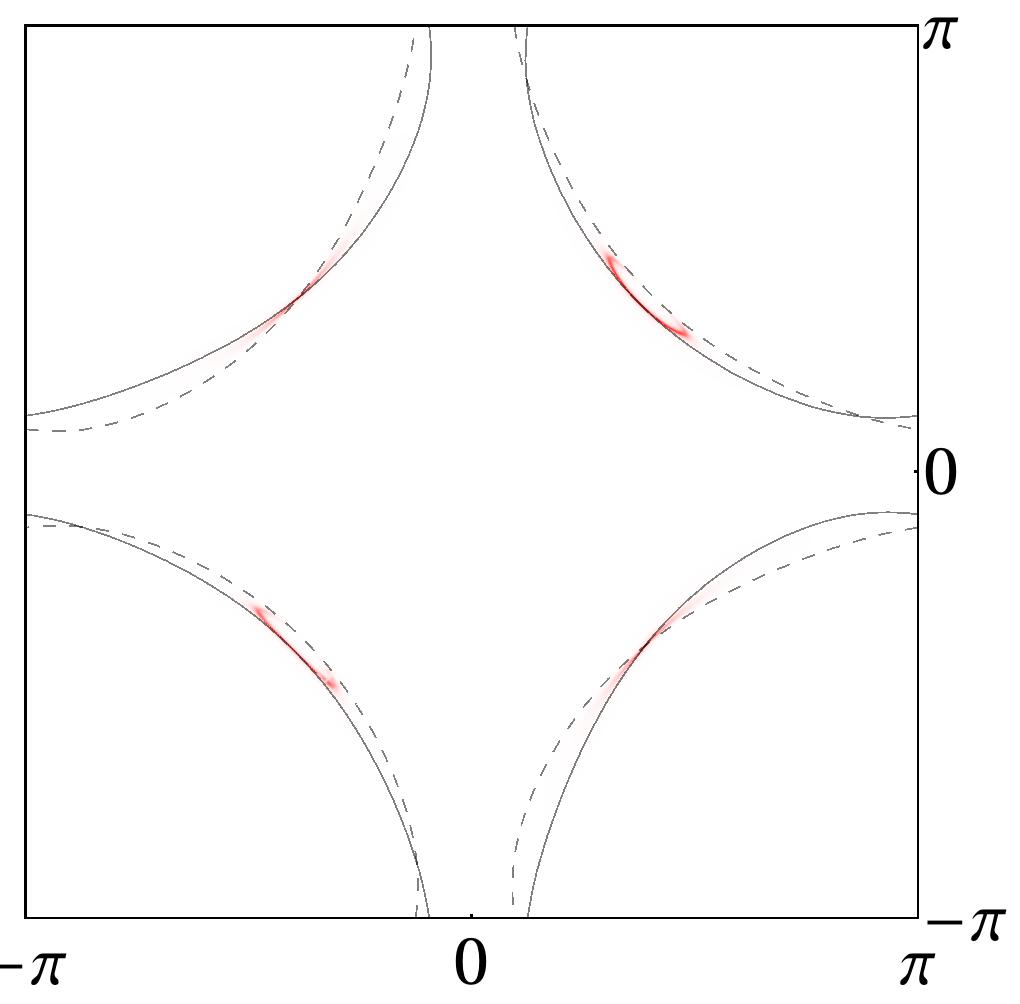}

\end{center}
\caption{\label{fig: gapless surfaces} $d$-wave superconductivity with loop current order. In solid black the normal state Fermi surface, and in dashed black its conjugate under $k \to -k$. On the left, blue regions have electron-like excitations, white regions have hole-like excitations, and the red line is the quasiparticle Fermi surface. On the right, the electron spectral function at the chemical potential. The upper plots have $v_\Delta < \lambda$, the lower plots have $v_\Delta > \lambda$. $\Delta_0 = 0.1$, $t_{1}=t_{2}=1$, $t_{4}=0.35$; $\mu=-1.18$, $t_{3}=-0.35-0.15i$ for the upper plots, $\mu=-1.2$, $t_{3}=-0.35-0.08i$ for the lower.}
\end{figure}

The presence of the extra term $\epsilon_{k}^{a}$, that would be forbidden by inversion/time reversal symmetries, allows the existence of gapless surfaces ($E^\pm_{k}=0$) in the superconducting state. This is illustrated in a typical example in fig. \ref{fig: gapless surfaces}.

Since the symmetry breaking is a small effect, it is reasonable to expand the quasiparticle energy about one of the four  nodal points $k_0$ where both $\epsilon^s_{k_0} = 0$ and \mbox{$\Delta_{k_0} = 0$}. There are four such points, as shown in fig. \ref{fig: gapless surfaces}. Let us write $k = k_0 + q$, and let us call $\hat{X}= \frac{\hat{x}+\hat{y}}{\sqrt{2}}$ (the nodal direction for points 1 and 2), $\hat{Y} =   \frac{\hat{x}-\hat{y}}{\sqrt{2}}$ the direction orthogonal to $X$. At point 1 and 2 we have, to first order in $|q|$
\begin{equation}\label{qp energy approx}
 E^\pm_{k_0 + q} = \epsilon^a_{k_0} \pm \sqrt{v_F^2 q_X^2 + v_\Delta^2 q_Y^2}\,,
\end{equation} 
where $v_F = |\nabla \epsilon^s_{k_0}|$, $ v_\Delta = |\nabla \Delta_{k_0}|$. More in general $v_\Delta$ and $v_F$ should be considered phenomenological parameters in a low energy theory of the system. The quasiparticle Fermi surface is given by the zeros of \eqref{qp energy approx}, and is an ellipse, within this approximation. At point 3 and 4 $\epsilon^a_{k_0} = 0 $\,,
so we have
\begin{equation}
 E^{\pm}_{k_0 + q} = \lambda q_Y \pm \sqrt{v_F^2 q_Y^2 + v_\Delta^2 q_X^2}\,,
\end{equation}
where $\lambda = |\nabla \epsilon^a_{k_0}|$. In this case\footnote{Before we disregarded the term proportional to $\lambda$ because it had to be compared with $v_F$, and could be neglected, under the hypotesis that the symmetry breaking is a small effect.}, if $v_\Delta > \lambda$, the $E_{\text{qp}}$ vanishes at the single point $k_0$, whereas, if $v_\Delta < \lambda$, the zeros of $E_{\text{qp}}$ are two lines that cross at $k_0$. From fig. 1 we see that they close and form a ``bowtie''.

Within this approximation, it is easy to compute the density of states at the chemical potential.  Further in the case (which we assume) $v_\Delta > \lambda$ this density of states comes entirely from the Fermi surfaces 1 and 2
\begin{equation}
 D^{FS} = 2 \int \ddbar^2\! k\, \delta(E_k) = \frac{\epsilon^a_{k_0}}{\pi v_F v_\Delta \hbar^2}\,,
\end{equation} 
The specific heat per mole for a n-layer cuprate material is then 
\begin{equation}
\begin{aligned}
 &C_v^{FS} = \gamma T\,,
 &&\gamma = \frac{\pi na^2\epsilon^a_{k_0} k_B R}{3 v_F v_\Delta \hbar^2}\,.
\end{aligned}
\end{equation} 

A residual $\gamma$ term in the specific heat is routinely measured in the superconducting cuprates at low-$T$ and is usually interpreted within dirty $d$-wave theory. However in YBCO the residual $\gamma$ is known to be roughly twice as large as in LSCO even though YBCO is much cleaner. Thus impurity effects on $d$-wave nodes may not entirely account for the observed $\gamma$ value. Another possibility is a contribution from localized chain electrons. If we take the loop order seriously there is a residual density of states coming from the Bogoliubov Fermi surface which also contributes to $\gamma$. An \emph{upper bound} on $\epsilon^a_{k_0}$ is obtained by attributing the full measured value $\gamma = 2 \text{mJ}/\text{mol}\, \text{K}^2$ in underdoped ortho-II YBCO \cite{riggs_heat_2011} to this contribution. Using the estimates $v_F \approx 1.8\, \text{eV}\,\text{\AA}$ and $v_\Delta \approx 0.1\, v_F$ we obtain (with $n = 2$ for YBCO) $\epsilon^a_{k_0} \lessapprox  30\,\text{meV}$.  We expect that a good fraction of the measured zero field $\gamma$ will come from the other two nodal points (within the usual dirty $d$-wave theory) and chain electrons so the actual value of $\epsilon^a_{k_0}$ will be a fraction of this upper bound.

From this upper bound we can also get the magnetic moment $M$ per triangle, using \eqref{current}.
\begin{equation}
 \frac{M}{\mu_B} = \frac{a^2 m_e}{\hbar^2} \langle j \rangle \lessapprox 0.02\,.
\end{equation} 
This is far smaller than the measured value $M/\mu_B \approx 0.1$ \cite{PhysRevB.78.020506}. However this may not be a very meaningful comparison. The measured moment points at an angle to the Cu-O plane and cannot be due to a pure orbital current that lives in the plane. Nevertheless it emphasizes the qualitative point that the large value of the moments reported in the experiments may lead to sizeable effects on the quasiparticle dispersion in the superconducting state which can then be looked for. 

It is instructive to calculate the electron spectral function in the SC coexisting with loop current order. 
\begin{equation}
\begin{split}
 A_k(\omega) = & \frac{1}{2}\left(1 + \frac{\epsilon^s_k}{E^0_k}\right) \delta(\omega - E^+_k)\, +\\
  &\frac{1}{2}\left(1 - \frac{\epsilon^s_k}{E^0_k}\right) \delta(\omega - E^-_k)
\end{split}
\end{equation} 
At zero frequency, it has weight only on the quasiparticle Fermi surface, and the spectral weight varies from $Z = 1$ at the nodal crossing of the normal state Fermi surface, to $Z = 0$ at the nodal crossing of its reflection conjugate. This is also illustrated in fig. \ref{fig: gapless surfaces}, right. In a real sample there would be domains realizing each inequivalent broken symmetry pattern, so real ARPES data would show a superposition with equal weight of all $4$ rotated and reflected images of fig. \ref{fig: gapless surfaces}, with or without ``bowties'' depending on $\lambda / v_\Delta$. For a scan along the nodal direction this will show up as two nodal quasiparticle peaks that are split in momentum by an amount $\Delta k = \epsilon^a_{k_0} / v_F$. Our earlier bound for $\epsilon^a_{k_0}$ then gives $\Delta k \lessapprox 0.02\, \text{\AA}^{-1}$. 
Interestingly a splitting of the nodal quasiparticle peak of roughly one third this magnitude was seen a number of years back in high resolution ARPES in underdoped Bi-2212 \cite{PhysRevB.66.014502,PhysRevB.75.140513}, and was interpreted as a bilayer splitting. Such bilayer splitting is a bit surprising due to the well known suppression of the c-axis hopping matrix element along the nodal direction. Our results obviously suggest an alternate interpretation in terms of the two pieces of the Bogoliuibov Fermi surface expected if loop order coexists with superconductivity. An interesting test of this interpretation will be to look for similar splitting in single layer cuprates, in particular Hg-1201 where loop order has also been reported in the normal state \cite{li_unusual_2008}.

\begin{figure}
\begin{center}
\includegraphics[width=0.5\columnwidth]{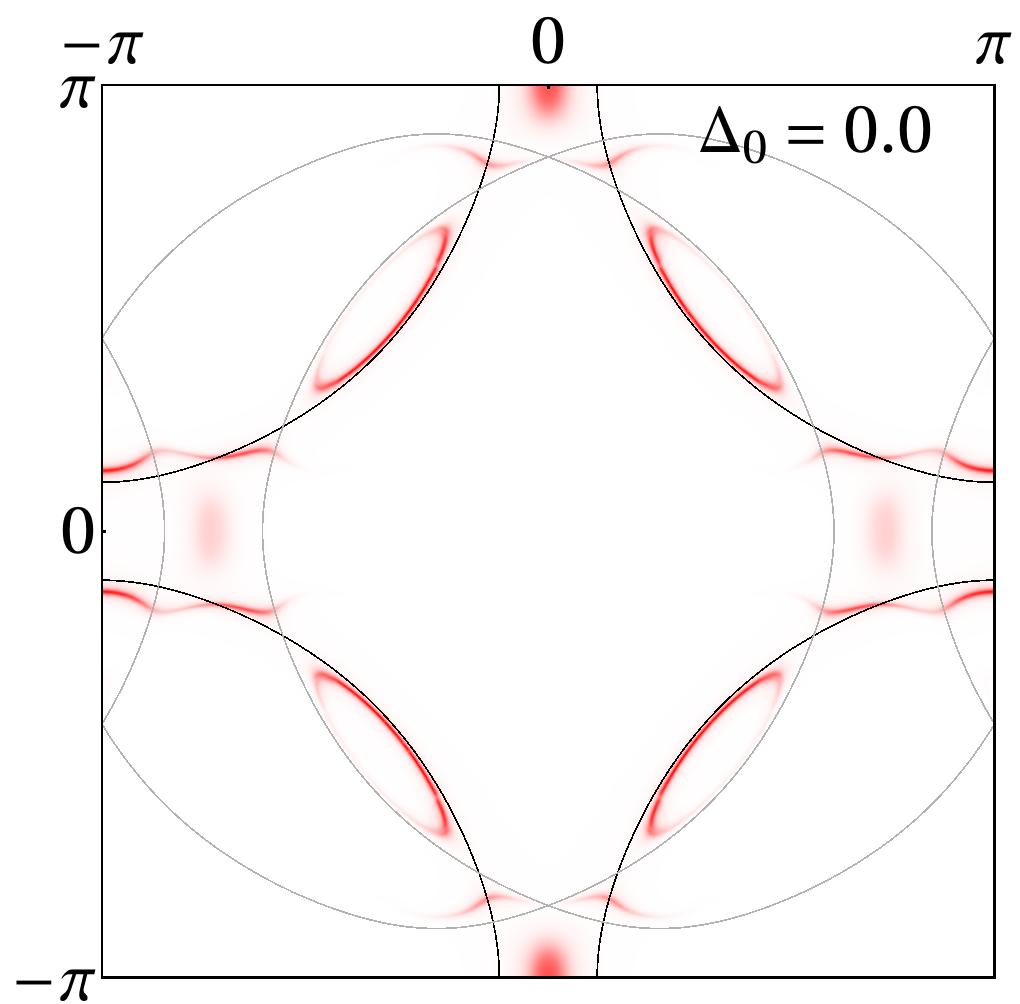}%
\includegraphics[width=0.5\columnwidth]{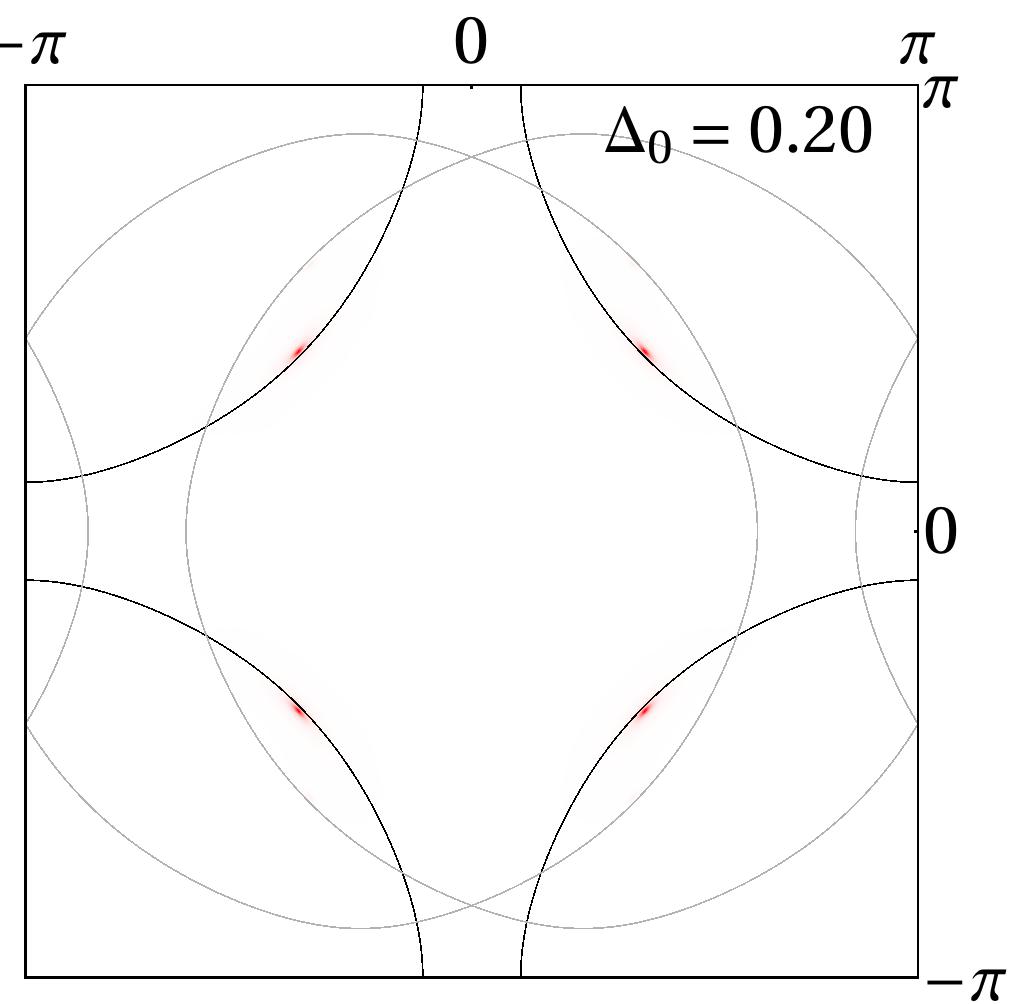}

\includegraphics[width=0.5\columnwidth]{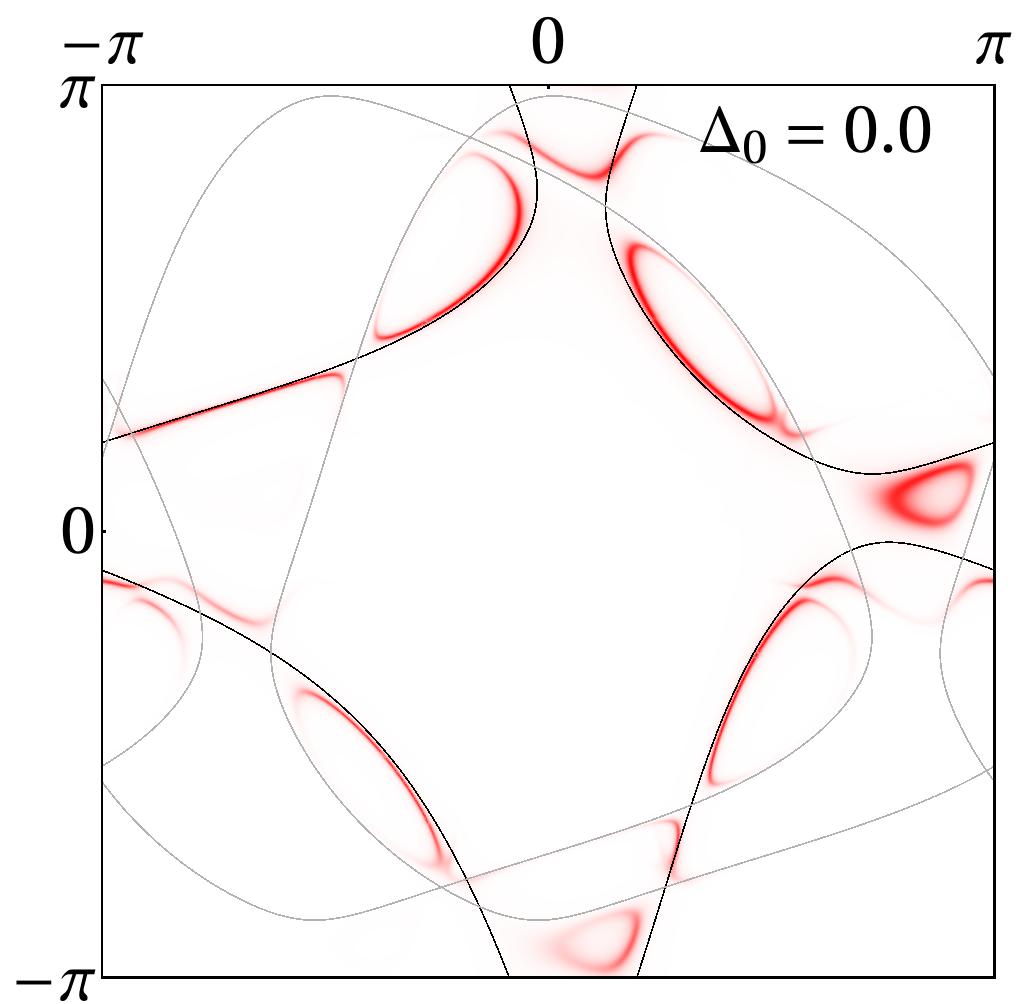}%
\includegraphics[width=0.5\columnwidth]{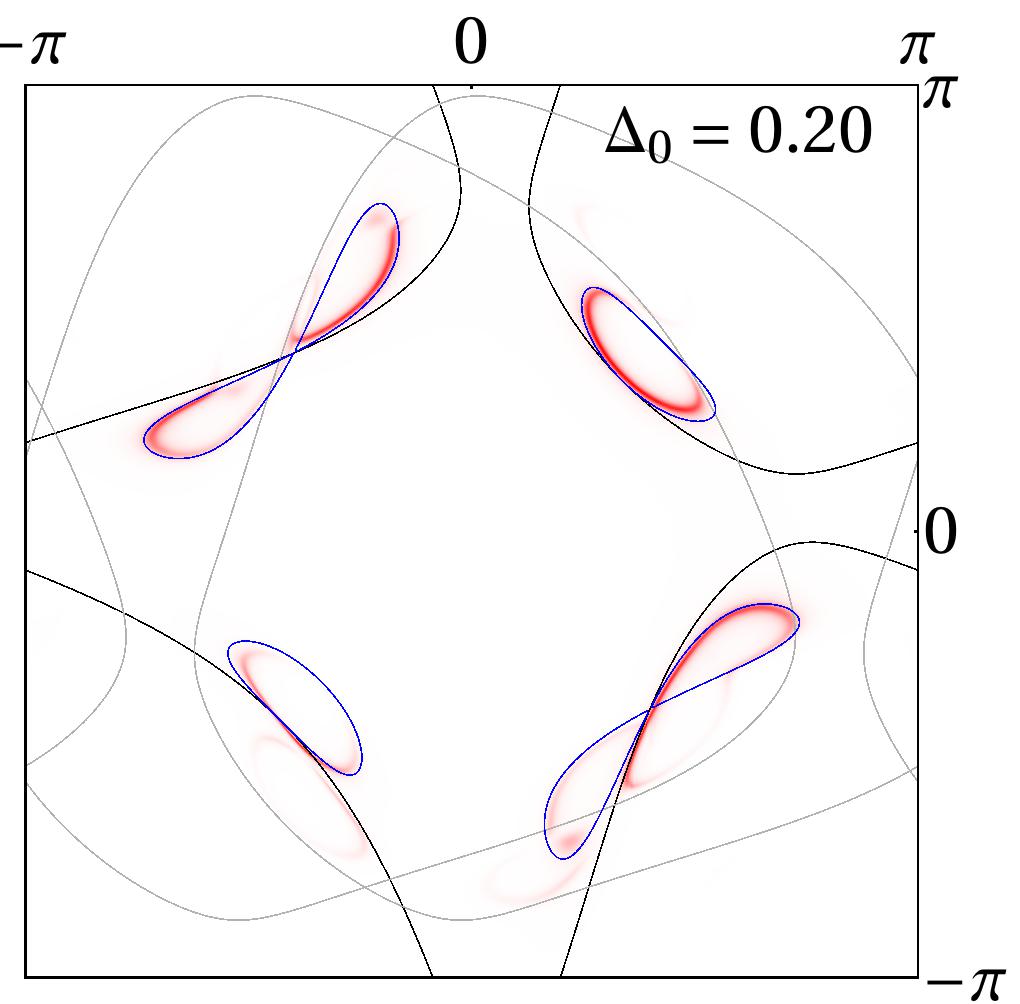}
\end{center}
\caption{\label{fig: survival of the pocket} Electron spectral function at the chemical potential. The upper plots do not have loop current order, the lower ones do. In all plots is present SDW order with period $Q = \left(3\pi / 4, \pi\right)$, amplitude $S = 0.2$. The plots on the right also have $d$-wave SC order. For the upper plots $\mu = - 1.2$, $t_{1}=t_{2}=1$, $t_{3} = t_4 = -0.35$. For the lower, $\mu = -1.11$, $t_3 = -0.35 - 0.3i$.}.
\end{figure}

\section{Effect of a magnetic field}
It is interesting to consider the behavior in a perpendicular magnetic field $H$. At low field the density of states of the Bogoliuibov Fermi surface will only have a weak dependence on $H$ while  the nodal quasiparticles associated with k-points $3,4$ will give the usual $\sqrt{H}$ behavior. Thus the full $\gamma$ coefficient in the specific heat will have a 
field dependent part that increases as $\sqrt{H}$. 
The presence of a gapless Bogoliubov Fermi surface immediately raises the question of whether a dSC coexisting with loop order will show quantum oscillations at low magnetic fields. Bogoliubov quasiparticles are superpositions of electron and hole and hence have indefinte charge. As we go around this Fermi surface the quasiparticles change in character from predominantly hole-like to predominantly electron-like, as displayed in Fig.~\ref{fig: gapless surfaces}.  Consequently, in a magnetic field, we do not expect the Bogoliubov Fermi surface to form Landau levels at the chemical potential and show quantum oscillations. We confirm this expectation with explicit calculations. 

In the cuprates, however, there is increasing evidence for the nucleation of translation symmetry breaking order (stripes) in a magnetic field. This is seen most directly in recent high field NMR experiments \cite{wu_magnetic-field-induced_2011} in the form of static period-4 charge order. Previously neutron experiments had suggested the presence of field-induced spin stripe order \cite{lake_spins_2001, PhysRevB.71.220508, PhysRevLett.102.177006, PhysRevLett.103.017001}. The NMR experiments do not detect any static spin stripe order but this may conceivably be due to the slower time scale of NMR as compared to neutron scattering. In any case a meaningful discussion of quantum oscillations in the cuprates must take into account the presence of field induced broken translation symmetry. We therefore consider the effects of stripe order on the `normal' state and  Bogoliubov Fermi surfaces. 

The most important observation is that if the normal state Fermi surface gets reconstructed by spin density wave order, then it can give rise to quantum oscillations even in the superconducting state, since, in some regions, it is protected from the SC gap.
This phenomenon is evident from fig. \ref{fig: survival of the pocket}. The plots on the left display Fermi pockets due to recostruction of the normal state Fermi surface from SDW order. The lower plots have loop current order, the upper do not. On the right, $d$-wave superconductivity is introduced. When there is no loop current order, the pockets get rapidly gapped, whereas, in presence of the loop current order, they are able to survive superconductivity.

\begin{figure}
\begin{center}
\includegraphics[width=\columnwidth]{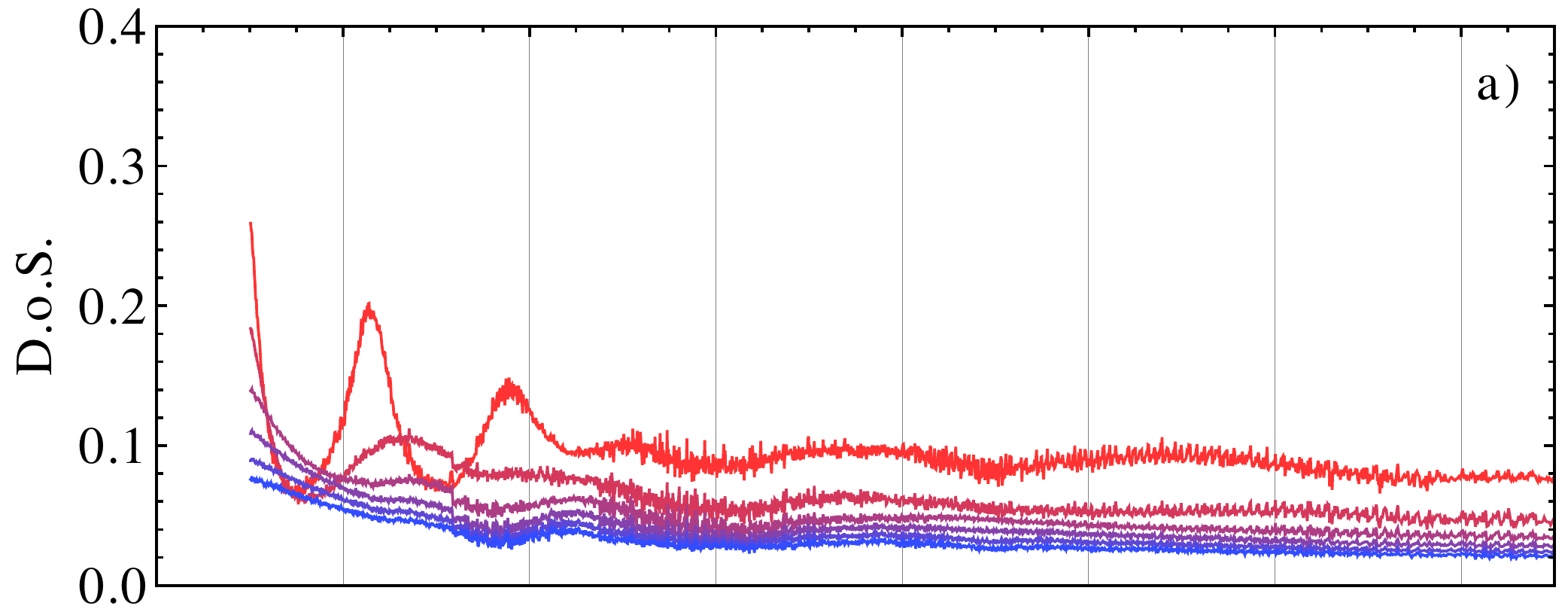}

\includegraphics[width=\columnwidth]{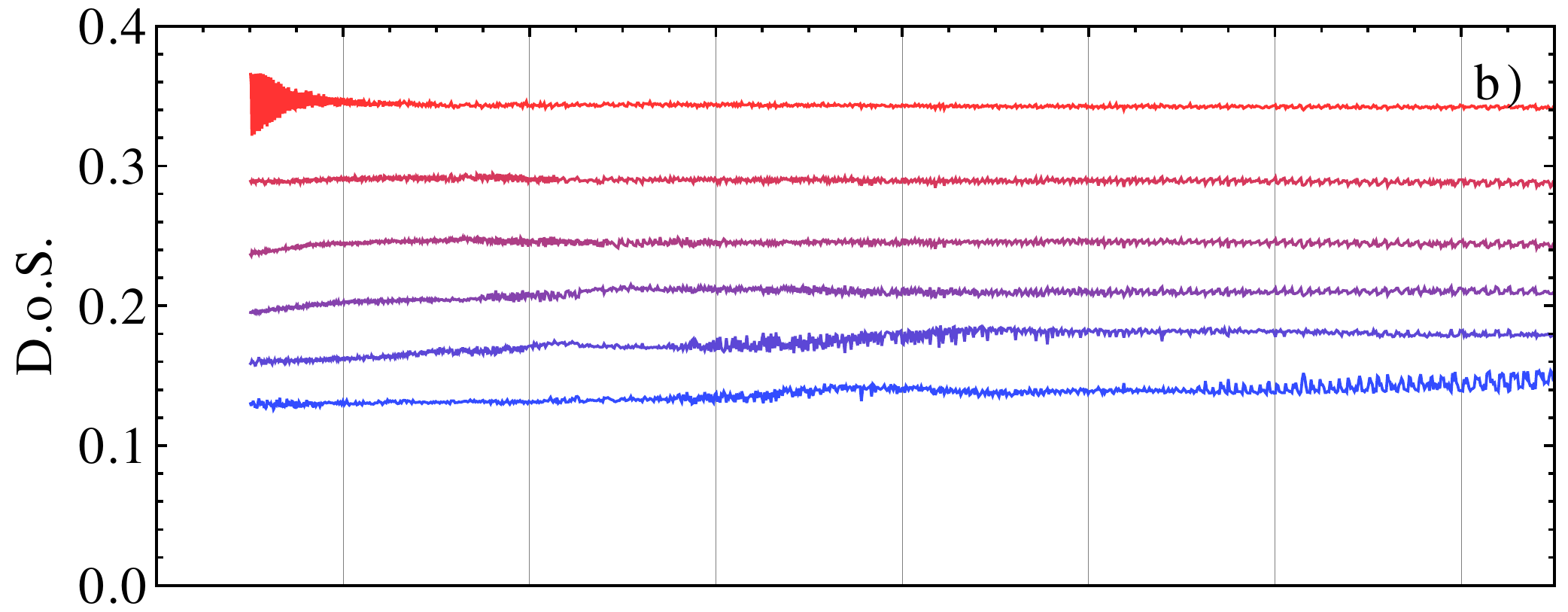}

\includegraphics[width=\columnwidth]{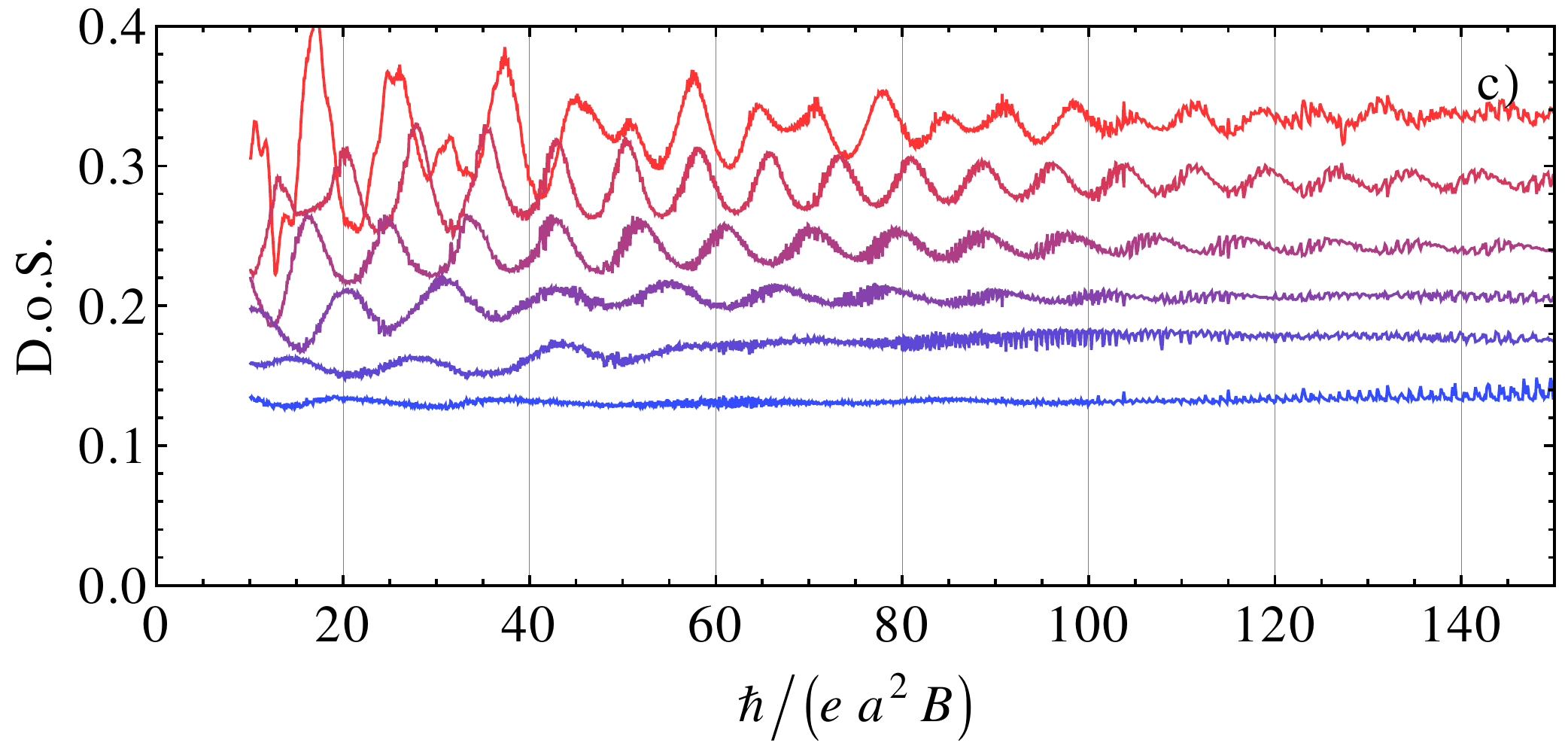}

\includegraphics[width=\columnwidth]{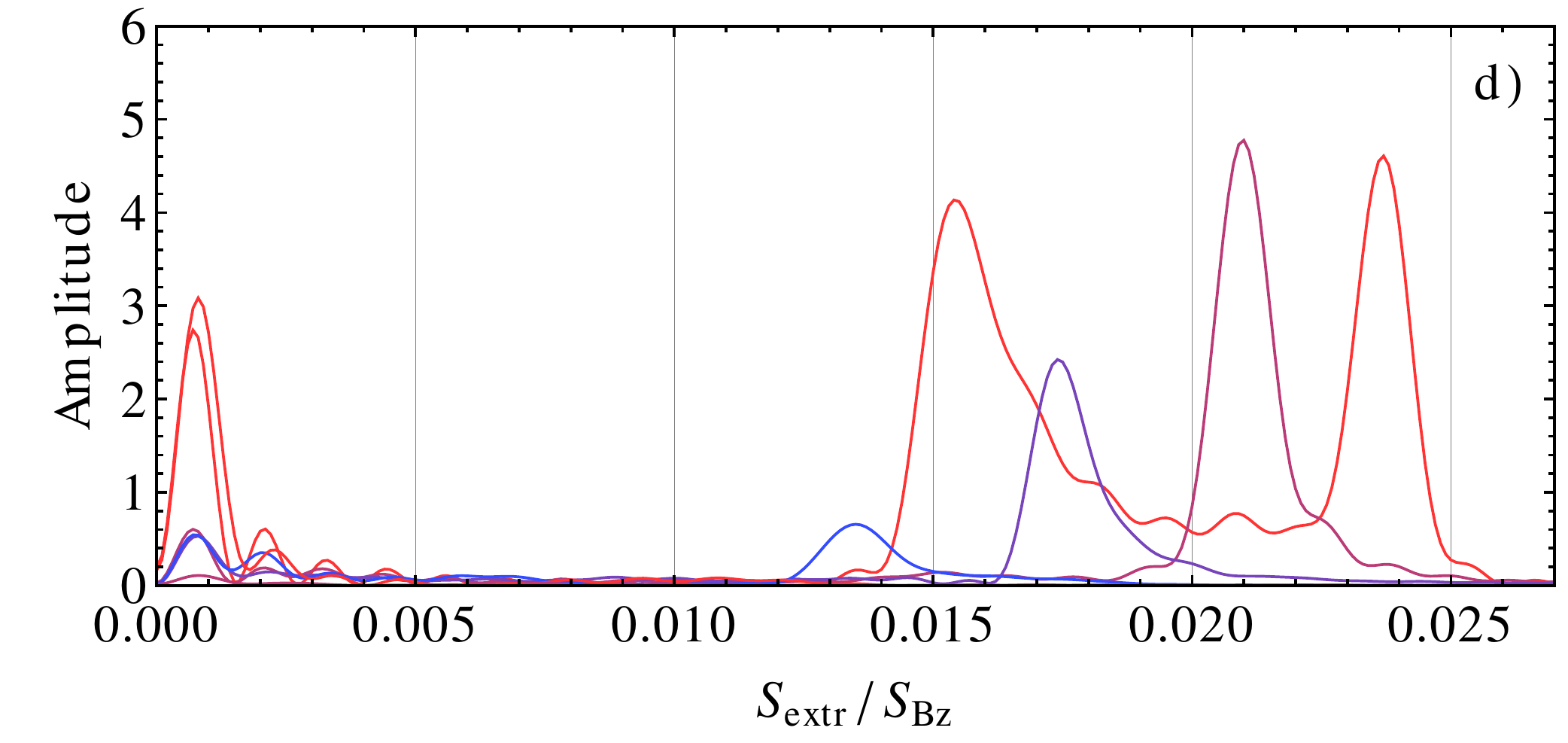}
\end{center}
\caption{\label{fig: oscillations} Density of states at the chemical potential as a function of $1/B$, for several values of $\Delta_0$. a) SDW order, no loop current order. b) loop current order, no SDW order. c) SDW and loop current order. d) Fourier transform of c), as a function of the orbit surface. The value of the superconducting order parameter increases from red to blue, and takes values $\Delta_0 = 0.05,\, 0.10,\, 0.15,\, 0.20,\, 0.25,\, 0.30$. $\mu = - 1.03$, $t_{1}=t_{2}=1$, $t_3 = -0.35 - 0.37i$, $t_4 = -0.35$.  The lattice has size $1000 \times 50$}
\end{figure}

To conclusively verify that the SDW pockets, protected by the loop current order, can cause quantum oscillations, we coupled the system to an external magnetic field. We use the transfer matrix method of Ref. \onlinecite{PhysRevB.79.180510} to directly calculate the density of states at the chemical potential. Our calculations are done on strips of size $50 \times 1000$.  Fig. \ref{fig: oscillations} shows a comparison between the systems with and without loop current order. In the first case, quantum oscillations are immediately killed by a very small SC order parameter, whereas, when loop order is present, they are able to survive up to modestly large values of $\Delta_0$. In this case, it is also noticeable that the period of the oscillations increases with $\Delta_0$. In general, the size of the orbit, as can be deduced from the oscillation period through the Onsager relation, is compatible with the size of the pocket as seen in the spectral function.

Though the above shows that a superconducting state with coexisting loop current  and spin stripe order can show quantum oscillations, there are a number of difficulties with postulating that this is what is actually going on the cuprate experiments. First, at least part of the high field data showing quantum oscillations is in the resistive (not superconducting) state. This may be dealt with by assuming that the high field state is a vortex liquid which (on the time scales of a cyclotron orbit) retains the electronic structure of the superconductor. Similarly, it is presumably enough that spin density wave order exist on the same time scales but need not be truly static.  The most serious difficulty is quantitative. The size of the orbit is more or less bounded by the size of the quasiparticle Fermi surface. In order to be able to observe oscillation compatible with a 2 \% orbit, as seen in experiments, a very large symmetry breaking is needed. For the plot in fig. \ref{fig: oscillations},  $\epsilon^a_{k_0} = 41\, \text{meV}$, assuming $t = 100\, \text{meV}$.
Such a large value of $\epsilon^a_{k_0}$ is in conflict with the upper bound discussed in the previous section coming from the measured zero field specific heat. Smaller values of 
$\epsilon^a_{k_0}$ consistent with the bound produce pockets that give far too small for they to be directly responsible for the observed oscillation phenomena. 

\section{Discussion}
We studied the effects of loop current order on the $d$-wave superconducting state and showed that these have several observable consequences. In light of the large moments reported in the experiments we may expect that its effects on the superconducting quasiparticles are large enough to be observed. The most important consequences are the 
presence of a Bogoliubov Fermi surface\cite{berg_stability_2007} leading to a residual density of states at the chemical potential and the presence of split nodal quasiparticle peaks in photoemission spectra. We obtained bounds on these effects from existing data on YBCO. It will be interesting to obtain similar bounds on Hg-1201 where loop order has also been reported by neutron experiments. As this material is a single layer cuprate photoemission evidence of nodal quasiparticle splitting will be particularly striking as there is no possible confusion with bilayer splitting. 

Though the neutron experiments find evidence for loop current order, such order has thus far not been seen in local probes such as NMR\cite{PhysRevLett.101.237001} or $\mu$SR\cite{PhysRevLett.101.017001}. Thus experimental detection of the Bogoliubov Fermi surface at low $T$ in the superconductor, though indirect, will be a striking confirmation of the presence of loop order in various cuprates.

We discussed the possibility of quantum oscillations aided by the presence of loop order in the  superconducting state. We showed that the Bogoliubov Fermi surface does not by itself have quantum oscillations but if it is reconstructed by spin stripe order it does. However for a quantitative comparison with the experiments we need to invoke a loop order
that is bigger than the bounds derived from zero field specific heat measurements.

Finally, we note that a residual Bogoliubov Fermi surface is also present if the superconductivity involves non-zero momentum pairing, as in FFLO states or the special case of the $\pi$-striped superconductor discussed for $\text{La}_{2-x}\text{Ba}_{x}\text{Cu}\text{O}_4$. The possibility of quantum oscillations in such a state has been studied recently by Zelli \emph{et al.} \cite{zelli_mixed_2011}.

We thank John Berlinsky, Tim Chen, Catherine Kallin, Patrick Lee, Subir Sachdev, L. Taillefer, and Chandra Varma  for useful discussions. TS was supported by NSF Grant DMR-1005434.

\bibliography{refs_1}

\section{Outline of the methods used}
\subsection{Quantum oscillations}

Our method follows closely \cite{PhysRevB.79.180510}. In order to write the BCS hamiltonian in a manifestly gauge invariant way, we introduce two vector fields $A$ and $v$, that live on the links of the lattice. The vector potential $A_\nu(x)$ is determined, up to gauge transformations, by its line integral
\begin{equation}\label{holonomy A}
  \sum_{(x,\nu) \in \partial S} A_\nu(x) = \frac{e B S}{\hbar}\,,
\end{equation}
where $S$ is any surface bounded by lattice links. The field $v_\nu(x)$ is the phase gradient of a gas of vortices, whose coordinates are fixed, and is determined, again up to gauge transformations, by
\begin{equation}\label{holonomy v}
  \sum_{(x,\nu) \in \partial S} v_\nu(x) = 2 \pi \times n_S\,,
\end{equation} 
where $n_S$ is the number of vortices inside $S$. Using $v$, we can also introduce a phase field
\begin{equation}
  e^{i\phi(x)} = \exp\left[i \sum_{y = x_0}^x v_\nu(y)\right]\,.
\end{equation} 
which is well defined, since the line integrals along different paths from $x_0$ to $x$ differ by $2\pi$. We can then write the hamiltonian as
\begin{equation} \label{hamiltonian magnetic field}
\begin{split}
 H = \sum_{x, \nu} \Big[& - t_\nu e^{i A_\nu(x)} \left(c^\dag_{x+ \nu \uparrow}c_{x\uparrow} +  c^\dag_{x +  \nu \downarrow}c_{x\downarrow}\right) - \\ &\Delta_\nu e^{i \phi(x) + \frac{i}{2} v_\nu(x)} c^\dag_{x+ \nu \uparrow}c^\dag_{x\downarrow} - \text{h.c.} + \\
 & S \cos Q \cdot x \left(c^\dag_{x\uparrow}c_{x\uparrow} - c^\dag_{x\downarrow}c_{x\downarrow}\right)\Big]\,,
\end{split}
\end{equation}

The hamiltonian is invariant under the following gauge transformation:
\begin{equation}
\left\{
\begin{aligned}
 &c_{x,\sigma} &&\to e^{i \theta(x)} c_{x,\sigma} \\
 &A_\nu(x) &&\to A_\nu(x) - \theta(x) + \theta(x + \nu)\\
 &v_\nu(x) &&\to v_\nu(x) - 2\theta(x) + 2\theta(x + \nu)\\
 &\phi(x) &&\to \phi + 2 \theta(x)\,,
\end{aligned}
\right.
\end{equation} 
so the hamiltonian effectively depends only on the gauge invariant field
\begin{equation}
 w_\nu(x) = v_\nu(x) - 2 A_\nu(x)\,,
\end{equation} 
and on the location of the vortices. In a variational approach, we should choose these quantities so as to minimize the ground state energy of \eqref{hamiltonian magnetic field}, compatibly with the constraints \eqref{holonomy A}, \eqref{holonomy v}. Since this is computationally unfeasible, in an effective approach, we determine the locations $r_i$ of the vortices by minimizing
\begin{equation}
 U_v = \sum_{i,j} \frac{1}{|r_i - r_j|}\,,
\end{equation} 
and we determine $w_\nu$ by minimizing
\begin{equation} \label{U w}
 U_w = \sum_{x, \nu} \left[v_\nu(x) - 2 A_\nu(x)\right]^2\,.
\end{equation} 
More precisely, we start with some choice of $A$ and $v$ that satisfy \eqref{holonomy A}, \eqref{holonomy v}, then we send $A_\nu(x) \to A_\nu(x) - \alpha(x) + \alpha(x + \nu)$, with $\alpha$ chosen so as to minimize \eqref{U w}. Since $U_w$ is quadratic in $\alpha$, we can minimize it by solving a linear system.

\begin{figure}
\begin{center}
\includegraphics[width=0.5\columnwidth]{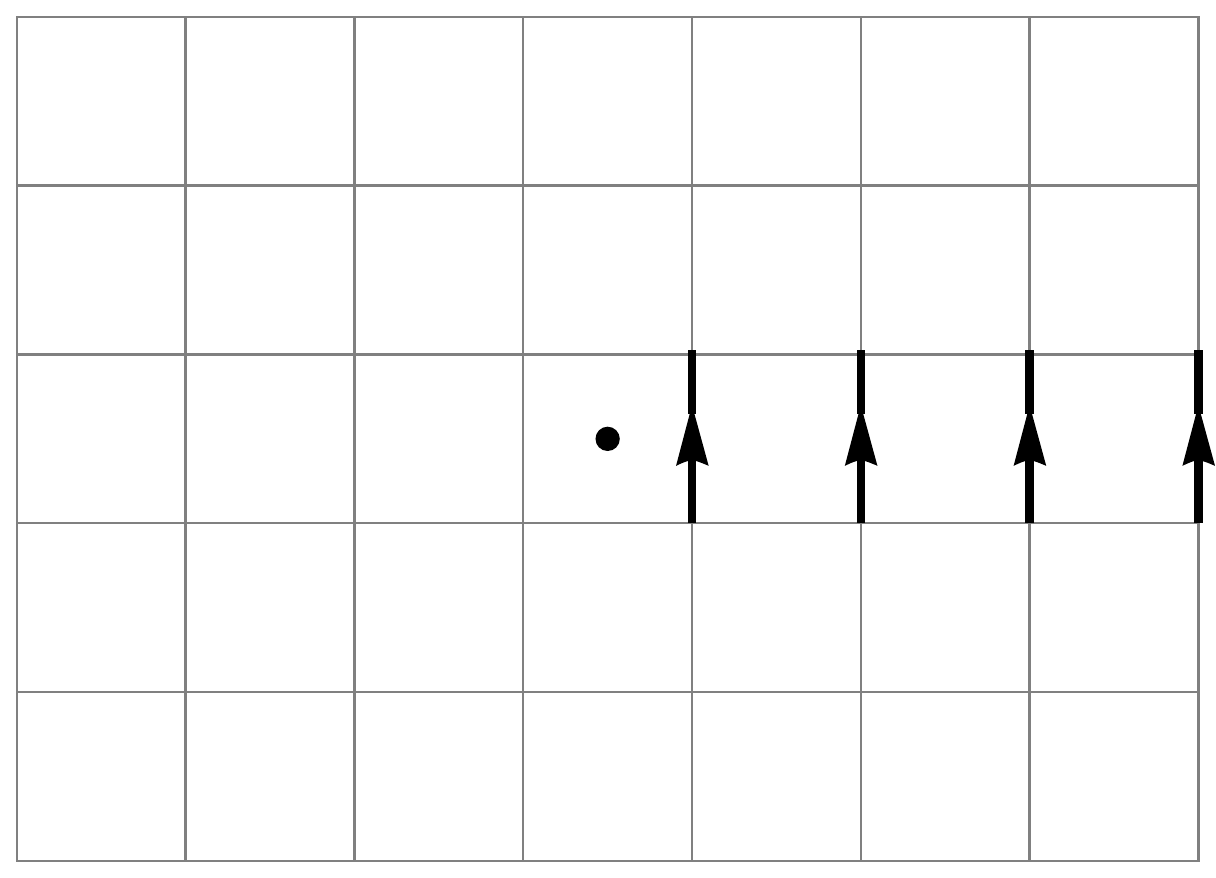}%
\includegraphics[width=0.5\columnwidth]{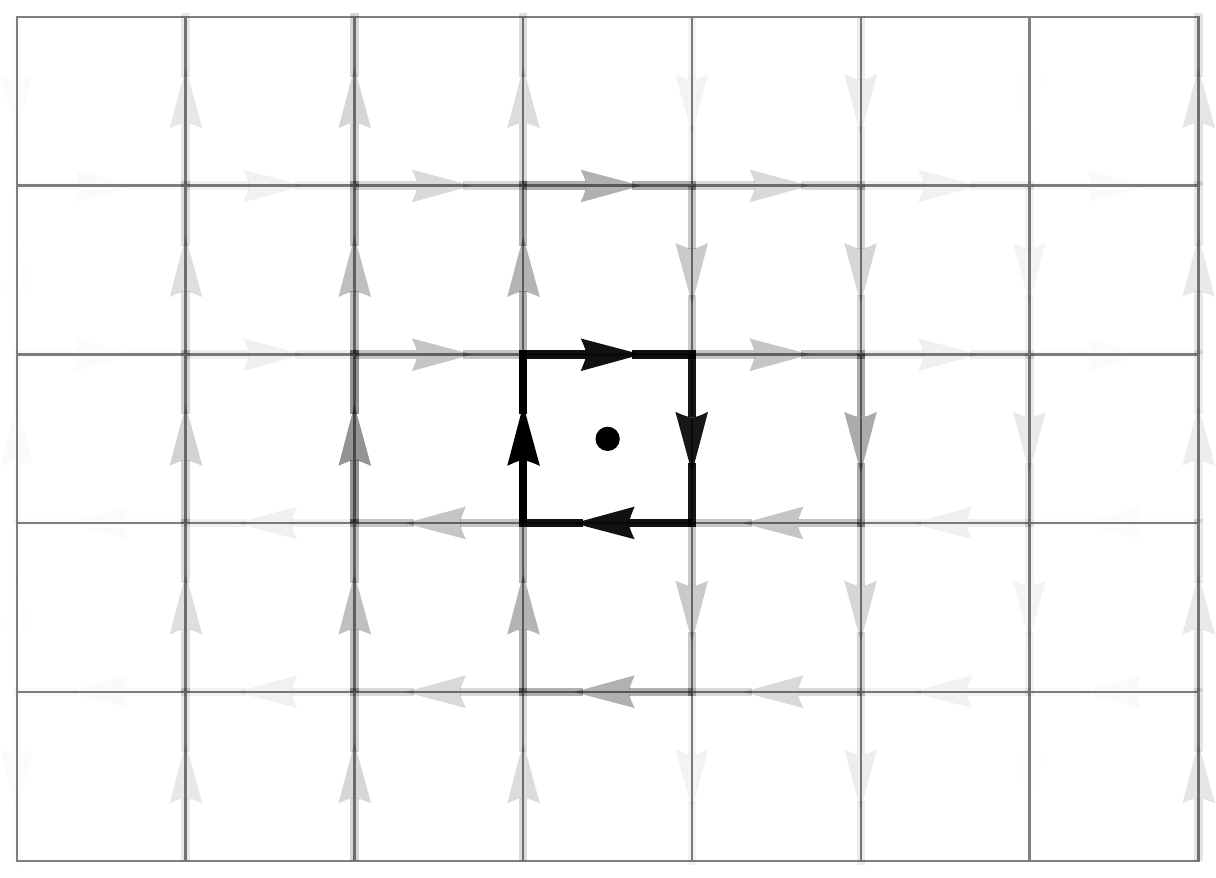}
\end{center}
\caption{Example of background fields configuration near a vortex. On the left, the field $v$. On the right, the field $w$ after energy minimization.}
\end{figure}

After performing the particle hole transformation
\begin{equation}
 c_{x, \uparrow} = d_{x, \uparrow}\,,
 \quad
 c_{x, \downarrow} = d^\dag_{x, \downarrow}\,,
\end{equation} 
the hamiltonian can be cast in the form
\begin{equation}\label{single particle hamiltonian}
 H = \sum_{(x_1 \sigma_1), (x_2\sigma_2)} h_{(x_1\sigma_1), (x_2\sigma_2)} d^\dag_{x_1\sigma_1} d_{x_2\sigma_2}\,,
\end{equation} 
so that the density of states can be written as
\begin{equation}
 D(\omega) = \mathrm{Tr}\left[\delta(h - \omega)\right] = \frac{1}{\pi} \mathrm{Im} \mathrm{Tr}\left[(h - \omega - i0^+)^{-1}\right]\,.
\end{equation} 

We choose periodic boundary conditions in the $y$ direction, with $N_y$ sites, and open boundary conditions in the $x$ direction, with $N_x$ sites. This restricts the allowed configurations of $v$ to those whose line integral along the $y$ direction is a multiple of $2 \pi$, so that the phase $\phi$ is well defined on the cylinder. With this choice of boundary conditions, if the indices $(x,y,\sigma)$ are ordered with the label $x$ changing slowest, the matrix $h$ is block-tridiagonal, with blocks of size $2N_y$, labelled by $x \in \{1, \ldots, N_x\}$:
\begin{equation}
 h = \begin{pmatrix}
     h_{11} & t_{12} & 0 & \ldots \\ 
     t_{21} & h_{22} & t_{23} & \ldots  \\
     0 & t_{32} & h_{22} & \ldots  \\
    \vdots & \vdots&\vdots
     \end{pmatrix}\,.
\end{equation} 

The diagonal blocks of $G = h^{-1}$ can be efficiently ($\text{time} \sim N_x N_y^3$) calculated with the following iterative algorithm
\begin{equation}
\begin{split}
&\begin{aligned}
 &L_1 = 0\,,
 &&
 L_{x+1} = t_{x+1,x}\left( h_{x,x} - L_x \right)^{-1} t_{x, x+1}\\
 & R_{N_x} = 0\,,
 &&
 R_{x-1} = t_{x-1,x}\left( h_{x,x} - R_x \right)^{-1} t_{x, x-1}
\end{aligned}\\
&\;G_{x,x} = \left( h_{x,x} - L_x - R_x \right)^{-1}\,.
\end{split}
\end{equation} 

\subsection{Spectral functions}
At zero magnetic field, the system is translationally invariant, and it is interesting to look at the electron spectral function
\begin{equation}\label{spectral function}
\begin{split}
 A_{k\sigma}(\omega) = &\bra{\text{gd}} c_{k\sigma} \delta(\omega - H) c^\dag_{k\sigma} \ket{\text{gd}}\\
& + \bra{\text{gd}} c^\dag_{k\sigma} \delta(\omega + H) c_{k\sigma} \ket{\text{gd}}\,,
\end{split}
\end{equation} 
where $H$ is the many body hamiltonian, and $\ket{\text{gd}}$ is its ground state, which is taken to have zero energy. With $H$ as in \eqref{hamiltonian magnetic field}, we can rewrite \eqref{spectral function} in terms of single particle quantities
\begin{equation}
\begin{aligned}
 &A_{k\uparrow}(\omega) = \bra{k,\uparrow} \delta(\omega - h) \ket{k,\uparrow}\\
 &A_{k\downarrow}(\omega) = \bra{-k,\downarrow} \delta(-\omega - h) \ket{-k,\downarrow}\,,
\end{aligned}
\end{equation} 
where $h$ is the same as in \eqref{single particle hamiltonian}, with the understanding
\begin{equation}\label{spectral function 1p}
\begin{aligned}
 &h_{(x_1\sigma_1), (x_2\sigma_2)} = \bra{x_1,\sigma_1} h \ket{x_2,\sigma_2}\,,\\
 &\ket{k, \sigma} = \sum_x e^{-i k x} \ket{x, \sigma}\,.
\end{aligned}
\end{equation} 

When there is no SDW order, we have
\begin{equation}\label{SC hamiltonian}
 h = \sum_k 
  \begin{bmatrix}\ket{k,\uparrow} \\ \ket{k,\downarrow}\end{bmatrix}
  \begin{bmatrix}
    \epsilon_k & \Delta_k \\
    \Delta_k^\star & -\epsilon_{-k}
  \end{bmatrix}
  \begin{bmatrix}
    \bra{k,\uparrow} & \bra{k,\downarrow}
  \end{bmatrix}
\end{equation} 

When there is no superconductivity\footnote{Here we suppress the spin index $\sigma$.}, but commensurate SDW order, so that $2Q = (2\pi m, 2\pi n) $, then
\begin{equation}\label{SDW hamiltonian 1}
 h = \sum_{k} 
  \begin{bmatrix}\ket{k} \\ \ket{k + Q}\end{bmatrix}
  \begin{bmatrix}
    \epsilon_k & S \\
    S & \epsilon_{k + Q}
  \end{bmatrix}
  \begin{bmatrix}
    \bra{k} & \bra{k + Q}
  \end{bmatrix}\,,
\end{equation}
where the sum runs over only half of the Brillouin zone. For higher order commensuration, $pQ = (2\pi m, 2\pi n)$, a $p\times p$ matrix can be introduced on the lines above, and the sum must run over a fraction $1/p$ of the brilluin zone. However, this approach becomes rapidly unfeasible as $p$ grows. A reasonable approximation is to introduce fictional copies
\begin{widetext}
\begin{equation}\label{SDW hamiltonian 2}
 h = \sum_{k} 
  \begin{bmatrix}\vdots \\ \ket{k, -1} \\ \ket{k} \\ \ket{k, +1} \\ \vdots\end{bmatrix}
  \begin{bmatrix}
    &\vdots & \vdots & \vdots &\\
    \hdots& \epsilon_{k - Q} & S & 0 &\hdots\\
    \hdots& S & \epsilon_k & S &\hdots\\
    \hdots& 0 & S & \epsilon_{k + Q}&\hdots\\
    &\vdots & \vdots & \vdots &\\
  \end{bmatrix}
  \begin{bmatrix}
    \hdots & \bra{k, - 1} & \bra{k} & \bra{k, + 1} & \hdots
  \end{bmatrix}\,,
\end{equation}
\end{widetext}
and have the sum run over the entire Brilloin zone. Usually one or two copies are sufficient. When both SC and SDW orders are present, the matrix structures of \eqref{SC hamiltonian} and \eqref{SDW hamiltonian 1} or \eqref{SDW hamiltonian 2} get combined in a straightforward way.

We evaluate \eqref{spectral function 1p} by diagonalizing these small matrices and smearing the delta function to a narrow Lorentzian.

For fig. \ref{fig: gapless surfaces} we defined ``particleness'' as
\begin{equation}
\begin{aligned}
 &|u_{k\uparrow}|^2 = \bra{k,\uparrow} \theta(-h) \ket{k, \uparrow}\\
 &|u_{k\downarrow}|^2 = \bra{-k,\downarrow} \theta(h) \ket{-k, \downarrow}\,.
\end{aligned}
\end{equation} 

\end{document}